\documentclass[10pt, conference]{IEEEtran}

\usepackage{amsmath}
\usepackage{graphicx}
\usepackage[colorlinks=true, allcolors=blue]{hyperref}
\usepackage{optidef}
\usepackage{amssymb}
\usepackage{bm}
\usepackage{subcaption}
\usepackage{svg}
\usepackage{placeins}
\usepackage{authblk}

\usepackage[style=acmnumeric]{biblatex}
\newcommand{\Var}{\mathrm{Var}}
\newcommand{\E}{\mathbb{E}}
\newcommand{\vect}[1]{\bm{#1}}

\title{Day-Ahead Bidding Strategies for Wind Farm Operators under a One-Price Balancing Scheme}

\author[1]{Max Bruninx}
\author[1]{Timothy Verstraeten}
\author[2]{Jalal Kazempour}
\author[1]{Jan Helsen}
\affil[1]{OWI-Lab - Vrije Universiteit Brussel, Brussels, Belgium} 
\affil[2]{Technical University of Denmark, Kgs. Lyngby, Denmark}

\begin{document}
\maketitle

\begin{abstract}
    We study day-ahead bidding strategies for wind farm operators under a one-price balancing scheme, prevalent in European electricity markets. In this setting, the profit-maximising strategy becomes an all-or-nothing strategy, aiming to take advantage of open positions in the balancing market. However, balancing prices are difficult, if not impossible, to forecast in the day-ahead stage and large open positions can affect the balancing price by changing the direction of the system imbalance. This paper addresses day-ahead bidding as a decision-making problem under uncertainty, with the objective of maximising the expected profit while reducing the imbalance risk related to the strategy. To this end, we develop a stochastic optimisation problem with explicit constraints on the positions in the balancing market, providing risk certificates, and derive an analytical solution to this problem. Moreover, we show how the price-impact of the trading strategy on the balancing market can be included in the ex-post evaluation. Using real data from the Belgian electricity market and an offshore wind farm in the North Sea, we demonstrate that the all-or-nothing strategy negatively impacts the balancing price, resulting in long-term losses for the wind farm. Our risk-constrained strategy, however, can still significantly enhance operational profit compared to traditional point-forecast bidding.
\end{abstract}

\section{Introduction}

The rise of zero-subsidy tenders in wind farm development increases the exposure of wind farms to electricity markets \cite{totalenergies2024negative, rvo2024negative}. Market exposure, however, poses an important challenge for wind farm operators, as the actual production and electricity prices are uncertain when submitting their bids to the day-ahead market. Additionally, deviations from the day-ahead schedule are settled by the transmission system operator in the balancing market and may result in imbalance penalties. The imbalance settlement procedure can either be \textit{two-price}, where the producer only incurs a penalty for positions that contribute to the overall system imbalance, or \textit{one-price}, where the producer is also rewarded for open positions that help the system to restore balance.  While the former provides an incentive to bid close to the point-forecast, the latter introduces arbitrage opportunities when taking a position in the market. 

In the case of a two-price balancing scheme, \cite{pinson2007trading} have shown that the problem can be rewritten as a Newsvendor problem, a famous problem in operations research, and solved analytically as an optimal quantile of the day-ahead production forecast. Many extensions to this work exist, such as the work by \cite{zugno2013trading}, where the optimal bid was constrained to avoid creating large imbalances, and the work by \cite{mazzi2016purely}, which showed that the optimal quantile solution can be learned from market data through a reinforcement learning algorithm. More recently, \cite{pinson2023distributionally} derived a distributionally robust version of the problem to deal with the fact that balancing prices are unknown a priori and difficult to forecast and \cite{munoz2023online} developed an online algorithm to apply the Newsvendor model in a non-stationary environment. 

In the past few years, however, the one-price balancing market became prevalent in European electricity markets \cite{acer2020harmon}. This shift from two-price to one-price balancing schemes, highlights an important gap in the literature, as most of the existing research focused on the day-ahead bidding problem within the context of a two-price scheme. The optimal bidding strategy under a one-price balancing scheme is fundamentally different, as it becomes an \textit{all-or-nothing strategy}, bidding either zero or the entire installed capacity to the day-ahead market. The former bid is placed when the expected day-ahead prices are smaller than the expected balancing prices, and the latter bid otherwise. However, the all-or-nothing strategy is a risk-seeking strategy that can result in serious financial losses during imbalance settlement. This can be caused by errors in price forecasts, which are difficult to estimate in the day-ahead stage, leading to wrong positions being taken in the market. Accurately forecasting balancing prices is particularly challenging because these prices depend on real-time system imbalances, which are often caused by systematic errors in day-ahead production forecasts of renewable energy generation. An alternative scenario occurs when the magnitude of the position is sufficiently large to change the direction of the system imbalance, causing the balancing price to shift from a reward to a penalty. To mitigate these risks, \cite{browell2018risk} proposed to constrain how much the bidding strategy can deviate from the point-forecast. While, the author shows that this is an effective method to reduce the risk related to the trading strategy, this approach does not consider the relationship between the uncertainty of the production forecasts and the balancing position explicitly. Moreover, the price-impact of the trading strategy on the balancing market was not considered during ex-post evaluation in their work. 

Our aim is to study the day-ahead bidding problem as a decision-making problem under uncertainty with the objective of maximising expected profit while reducing the imbalance risk related to open positions in the balancing market. To this end, we introduce a stochastic optimisation problem with explicit constraints on these positions, offering a \textit{risk certificate}, and derive an analytical solution to this problem. Using this solution, the optimal bidding strategy can be derived from forecasts of wind power generation and electricity prices. This enables our method to leverage the accuracy of data-driven forecasting techniques, while the actual decision-making process is based on an analytical formula, making it fully transparent and interpretable. Furthermore, we show how trading strategies can be evaluated while taking the price-impact on the balancing market of the strategy into consideration. To this end, historical balancing volume bids are used to evaluate ex-post how the trading strategy would have influenced balancing prices. To the best of our knowledge, this is the first paper that includes a realistic assessment of the price-impact on the balancing market in this context.

To evaluate our method, trading strategies are derived, ranging from point-forecast bidding to the all-or-nothing strategy, by varying the size of the constraint on the position in the balancing market. Their performance is evaluated with data from the Belgian electricity markets and an offshore wind farm in the North Sea during the first semester of 2024. The findings indicate that the trading strategy of a large-scale offshore wind farm can have an adverse impact on the balancing price, in case the imbalance positions become too large, resulting in long-term financial losses for the wind farm. Nonetheless, our risk-constrained strategy can still significantly enhance operational profit compared to traditional point-forecast bidding. Moreover, the results show how the profit distributions substantially undervalue the risk related to the trading strategy when the price-impact on the balancing market is not considered. This emphasizes the importance of accurate electricity market models when evaluating trading strategies on historical data. 

The remainder of the paper is structured as follows: Section \ref{sec:problem} formally describes the problem setting, outlining key details and assumptions, and defines a stochastic optimisation problem for day-ahead bidding. Section \ref{sec:analytical} presents an analytical solution to the problem. Section \ref{sec:experiments} focuses on the experimental results to validate our method. Finally, Section \ref{sec:conclusion} concludes the work and outlines a potential pursuit for future research. 

\textit{Note}: Throughout the paper, bold symbols denote vectors, capital calligraphic letters are used for sets, and capital Latin or Greek letters represent random variables. We use $\E[X]$ to denote the expected value of random variable $X$ and $\Var(X)$ to denote the variance. We use the $\tilde{(.)}$ to indicate that values are normalised.

\section{Problem statement}
\label{sec:problem}

\subsection{Overview}

We study the day-ahead bidding problem for a wind farm operator under a one-price balancing market. Every day, the wind farm operator needs to decide how much electricity to sell for each time unit, typically every hour, of the next day. In real-time, the difference between the actual production and the contracted volumes, i.e., the open positions in the balancing market, are settled against the balancing price. Given that we consider a one-price balancing market, imbalance settlement can either result in additional revenue or an imbalance penalty. The wind farm is rewarded for positions which help the system to restore balance, and penalised otherwise. At the day-ahead stage, actual production and electricity prices are unknown, making this a decision-making problem under uncertainty. To determine the optimal bidding strategy, we develop a stochastic optimisation problem where the objective is to maximise expected profit while constraining the size of the position in the balancing market. Compared to data-driven methods, stochastic optimisation allows us to explicitly control the objective and constraints of the optimisation problem, making the method more white-box.

\subsection{Notation} 

On the day-ahead market, contracts are sold over a time window denoted by $w_t^{\textrm{DA}} = \left[t, t + r^{\textrm{DA}} \right[ \in \mathcal{W}^{\textrm{\textrm{DA}}}$, where $r^{\textrm{DA}}$ is the time resolution of the tradable contracts and $\mathcal{W}^{\textrm{DA}}$ represents all contracts sold for a specific day. In most EU countries, the day-ahead auction closes at noon and tradable contracts have a 1-hour resolution \cite{epexspot2022trading}. In real-time, open positions in the balancing market are settled over a time window denoted by $w_t^{\textrm{B}} = \left[t, t + r^{\textrm{B}} \right[ \in \mathcal{W}^{\textrm{B}}$, with $r^{\textrm{B}}$ the time resolution of the imbalance settlement and $\mathcal{W}^{\textrm{B}}$ represents the imbalance settlement periods during the same trading day. The time resolutions of the day-ahead and balancing market are not necessarily equivalent. For example, imbalance settlement is presently carried out with a 15-minute resolution in Belgium \cite{eliaimbalance} and with a 1-hour resolution in the Nordic countries \cite{energinet2024imbalance}, whereas both countries operate a day-ahead market with a 1-hour resolution \cite{epexspot2022trading}. However, to simplify the problem formulation, the resolution of the day-ahead market and balancing market are considered to be equal $\mathcal{W}^{\textrm{DA}}=\mathcal{W}^{\textrm{B}}$ in Section \ref{sec:problem} and Section \ref{sec:analytical}. The day-ahead and balancing prices are, respectively, given by $\vect{\Lambda^{\textrm{DA}}} = \left[\Lambda^{\textrm{DA}}_t \: | \: \forall t: w^{\textrm{DA}}_t \in \mathcal{W}^{\textrm{DA}}\right]$ and $\vect{\Lambda^{\textrm{B}}} = \left[\Lambda^{\textrm{B}}_t \: | \: \forall t: w^{\textrm{B}}_t \in \mathcal{W}^{\textrm{B}}\right]$. The energy production is given by $\vect{E} = \left[E_t \: | \: \forall t: w^{\textrm{B}}_t \in \mathcal{W}^{\textrm{B}}\right]$. Note that the actual electricity prices and production are unknown at the moment of decision making, but forecasts of the statistical measures of these variables are available. The bidding strategy $\vect{y} = \left[y_t \: | \: \forall t: w^{\textrm{DA}}_t \in \mathcal{W}^{\textrm{DA}}\right]$ corresponds to the quantities of energy offered in the day-ahead market.  The installed capacity of the wind farm is given by $\beta$ and provides an upper bound on the actual production. Throughout the paper, installed capacity is in MW, energy production is in MWh and electricity prices are in €/MWh.

\subsection{Assumptions}

Three important assumptions are made to model the day-ahead bidding problem: 

First, the wind farm is assumed to be a \textit{price-taker}, meaning that the wind farm operator does not account for the potential impact of its bidding strategy on day-ahead and balancing market prices, nor does it leverage these effects to its potential advantage. However, in the balancing market, an offshore wind farm can potentially have price-impact when the open position is large compared to the system imbalance. As the installed capacity of offshore wind farms has grown over time \cite{bilgili2022global}, the likelihood of such a scenario has risen. The potential price-impact in the balancing market is, therefore, explicitly validated in our experiments. 

Second, we assume that \textit{offered quantities on the day-ahead market are always included in the market clearing}. This assumption was often made in past research \cite{pinson2007trading,rahimiyan2015strategic,browell2018risk,pinson2023distributionally} and is based on the fact that a renewable energy producer faces no marginal operational costs and can, therefore, offer at price zero. However, due to large renewable energy penetration, it is possible that the offer of the wind farm operator is not included in the market clearing or has a downward impact on market prices \cite{de2015negative}. Therefore, an interesting pursuit for future research would be to assess the impact of the day-ahead market clearing on the experiments.

Last, we assume that it is \textit{not possible to control the energy production of the wind farm} through curtailments or electrical energy storage. While, in practice, the energy production of a wind farm can be controlled through curtailments, this assumption allows us to focus on the bidding problem in this work. Nonetheless, we outline how the problem can be extended to include the possibility of curtailments in Section \ref{subsec:opti}. Additionally, this assumption implies that it is not possible to carry-over electricity production from one time unit to another, meaning that the optimal bidding strategy can be found for each tradable contract independently. 

\subsection{Optimisation problem}\label{subsec:opti}

We formulate the day-ahead bidding problem as a stochastic optimisation problem, where for each tradable contract $w_t\in\mathcal{W}$ we aim to find the optimal bidding quantity $y^*_t$ which:
\begin{maxi!}|s|
    {y_t}{\E\left[\Lambda_t^{\textrm{DA}} y_t+\Lambda^{\textrm{B}}_t\left(E_t-y_t\right)\right]\label{maxi:objective}}{\label{maxi:full_problem}}
    {}{}
    \addConstraint{\E\left[\left(E_t - y_t\right)^2\right]}{\leq\alpha}\label{maxi:const-imbalance}
    \addConstraint{0\leq y_t}{\leq\beta}\label{maxi:const-bid}.
\end{maxi!}

The objective function (\ref{maxi:objective}) corresponds to the expected profit resulting from the day-ahead and the balancing market. In the day-ahead market, the wind farm receives the day-ahead price for each contracted volume:
\begin{equation}
    \Lambda_t^{\textrm{DA}} y_t.
\end{equation}

The payoff from the balancing market, which can be positive or negative, is given by:
\begin{equation}
    \Lambda^{\textrm{B}}_t\left(E_t-y_t\right),
\end{equation}
where $\left(E_t-y_t\right)$ corresponds to the open position in the balancing market. Note that, in our case the balancing price is the same for every market participant, whereas in a two-price balancing market the balancing price would have been conditional on the direction of the open position and whether it aligns with the direction of the system imbalance or not.

Constraint (\ref{maxi:const-imbalance}) is introduced to reduce the expected risk related to open positions in the balancing market. The constant $\alpha$ within this constraint, referred to as the \textit{risk certificate}, represents the trade-off between risk and return. The squared difference $\left(E_t-y_t\right)^2$ is used to assess the imbalance risk of the trading strategy. This measure imposes a symmetric constraint on production surplus and deficit, increasing more strongly when the open positions becomes larger. To evaluate the imbalance risk, the squared difference is a more effective measure than the open position $\left(E_t-y_t\right)$, where production surpluses and deficits would cancel each other out in the calculation of the expected value. Constraint (\ref{maxi:const-bid}) enforces the bidding quantity to be between zero and the installed capacity of the farm.

The problem can be formulated more generally by writing constraint (\ref{maxi:const-imbalance}) as:
\begin{equation}
\E\left[f\left(E_t, y_t\right)\right]\leq\alpha,
\label{eq:gen_imb}
\end{equation}
with $f\left(.\right)$ a function which represents the imbalance risk. For instance, to include the possibility of curtailments, it can be defined as:
\begin{equation}
    f\left(E_t,y_t\right) = 
    \begin{cases}
        \gamma \left(E_t-y_t\right) & \text{if } E_t>y_t\\
        (1-\gamma) \left(y_t-E_t\right) & \text{else},
    \end{cases}
\label{eq:curtailments}
\end{equation}
where the constant $\gamma<0.5$ assigns a lower level of risk to a production surplus than to a production deficit. For example, $\gamma=0$ implies zero imbalance risk associated with production surplus, which requires that the wind farm can always curtail downwards up to the desired production level and that the balancing prices are known in real-time. However, in practice, these requirements are not always met, as curtailments affect the lifetime of wind turbines \cite{robbelein2023effect} and balancing prices are only known a posteriori \cite{pavirani2025predicting}. Therefore we recommend setting the constant $\gamma>0$. The complexity of the function also necessitates the use of integration or simulation to calculate the expected value, which are beyond the scope of this paper. Nonetheless, the application of the proposed methodology using (\ref{eq:curtailments}) offers an interesting direction for future research.

\section{Analytical solution}
\label{sec:analytical}

The optimal solution to problem (\ref{maxi:full_problem}) can be derived analytically. To illustrate this, we first derive the solution for the problem setting without the risk constraint (\ref{maxi:const-imbalance}) and show that it results in an all-or-nothing strategy. Then, we illustrate how this can be extended to our problem setting. The advantage of deriving an analytical solution is that it is computationally efficient and intuitive. The solution is derived under the same resolution for day-ahead and balancing markets. The extension of this solution to different resolutions can be found in Appendix \ref{app:full_problem}.

First, the objective function (\ref{maxi:objective}) can be rewritten as:
\begin{equation}
    \left(\E\left[\Lambda^{\textrm{DA}}_t\right] - \E\left[\Lambda^{\textrm{B}}_t\right]\right) y_t + \E\left[\Lambda^{\textrm{B}}_t E_t\right],
\end{equation}
where $\E\left[\Lambda^{\textrm{B}}_t E_t\right]$ is a constant term that does not impact the optimal solution, so it can be removed. This shows that the objective function is linear in $y_t$, implying that the optimal bidding quantity $y^{*}_t$ is given by:
\begin{equation}
    y^{*}_t = 
    \begin{cases}
        \beta & \text{if } \E\left[\Lambda^{\textrm{DA}}_t\right]>\E\left[\Lambda^{\textrm{B}}_t\right]\\
        0 & \text{else}.
    \end{cases}
\label{eq:all-or-nothing}
\end{equation}

Consequently, the optimal strategy becomes an all-or-nothing strategy, where the operator would bid the installed capacity when the expected day-ahead price is higher than the expected balancing price, and bid zero otherwise. To derive the optimal solution for the complete problem, the risk constraint (\ref{maxi:const-imbalance}) is first rewritten as:
\begin{equation}
     \E\left[E_t\right] - \sqrt{\alpha - \Var\left(E_t\right)} \leq y_t \leq \E\left[E_t\right] + \sqrt{\alpha - \Var\left(E_t\right)},
\label{eq:imb-const-rewrit}
\end{equation}
where we use the decomposition of the variance:
\begin{equation}
    \Var\left(X\right) = E\left[X^2\right] - E\left[X\right]^2.
\end{equation}

\begin{figure}[t!]
\centering
\includegraphics[width=\linewidth]{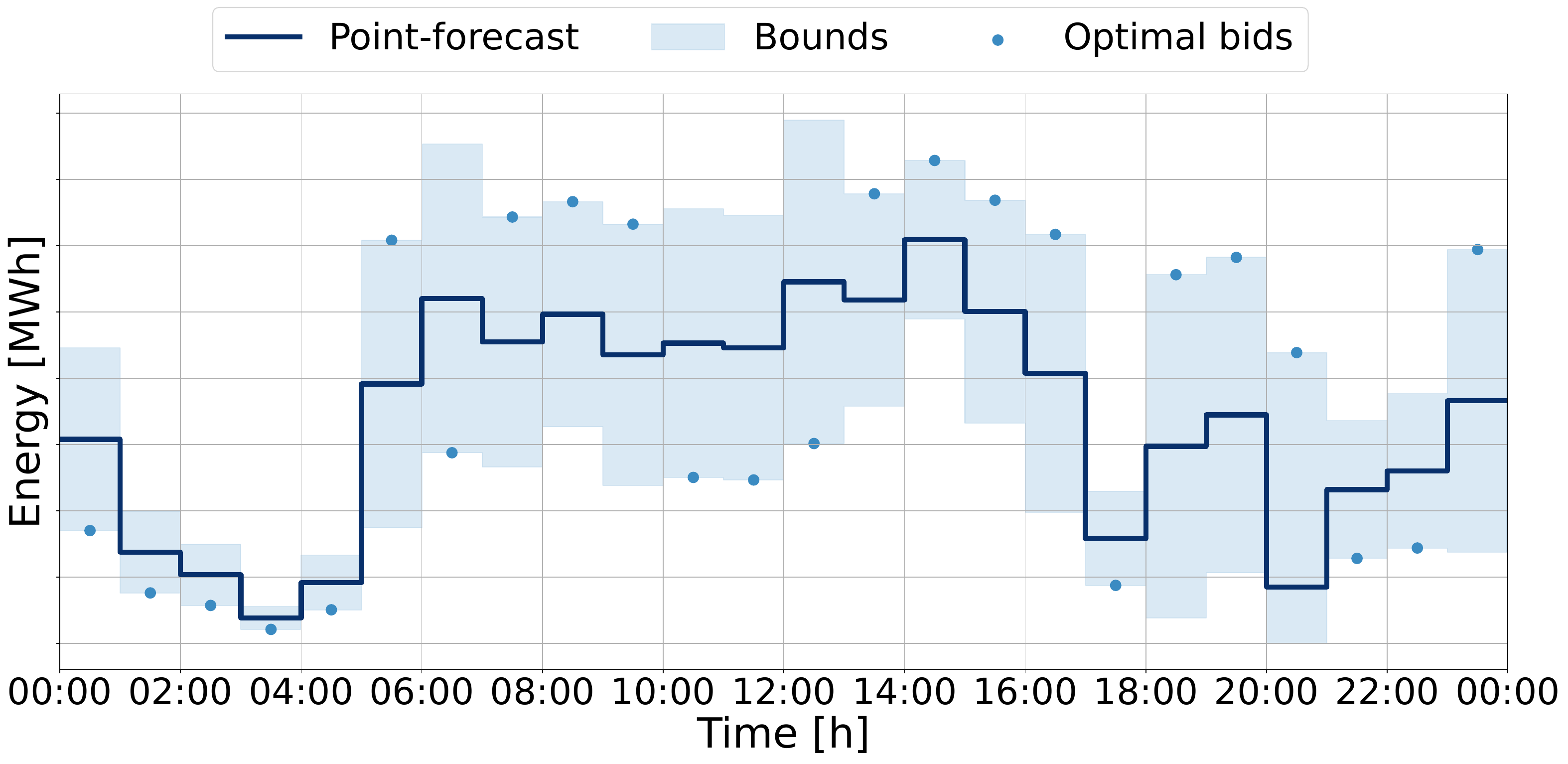}
\caption{Illustrative example of optimal day-ahead bidding strategy for a specific day. The bounds around the point-forecast are determined by risk certificate $\alpha$. The bidding strategy is long (e.g., in hour 00:00) when the expected day-ahead price is lower than the expected balancing price, leading to a bid below the production forecast. In contrast, the strategy is short (e.g., in hour 16:00) when the expected day-ahead price exceeds the expected balancing price, resulting in a bid above the production forecast.}
\label{fig:risk_bound}
\end{figure}

By substituting (\ref{maxi:const-imbalance}) with (\ref{eq:imb-const-rewrit}), and following the same argumentation used to obtain (\ref{eq:all-or-nothing}), the solution becomes:
\begin{equation}\label{eq:analytical_solution}
    y^{*}_t = 
    \begin{cases}
        \min\{\E[E_t]+\Delta_t ; \beta\} & \text{if } \E\left[\Lambda^{\textrm{DA}}_t\right]>\E\left[\Lambda^{\textrm{B}}_t\right]\\
        \max\{\E[E_t]-\Delta_t ; 0\} & \text{else}
    \end{cases},
\end{equation}
with $\Delta_t = \sqrt{\alpha-\Var(E_t)}$. Therefore, optimal bidding strategy becomes a binary decision where the trading position is bounded by an allowed deviation from the point-forecast. Figure \ref{fig:risk_bound} provides an illustrative example of how the trading strategy can be interpreted visually. This also illustrates how our work is related to the work by \cite{browell2018risk}. 

Note that the risk certificate $\alpha$ from constraint (\ref{maxi:const-imbalance}) is an important input parameter to our method which determines the risk-return trade-off of the trading strategy. Equation (\ref{eq:analytical_solution}) indicates how this parameter impacts the strategy. If the risk certificate $\alpha$ is set equal to the forecasted variance of the production $\Var\left(E_t\right)$, the optimisation problem reduces to minimising the expected imbalance and its solution is given by bidding the point-forecast $\E\left[E_t\right]$. If the risk certificate $\alpha$ becomes sufficiently large, the imbalance constraint will no longer be a binding constraint and the trading strategy becomes an all-or-nothing strategy. For any risk certificate in between those values, the trading strategy becomes a binary decision bounded by an allowed deviation $\Delta_t$ from the point-forecast.

\section{Experiments}
\label{sec:experiments}

\subsection{Set-up}

We evaluate our method out-of-sample using data of the first semester of 2024. To this end, we construct different trading strategies, based on the value of the risk certificate, and evaluate their cumulative performance over time. In addition, we study the profit distributions to assess the short-term financial risk of each trading strategy. Instead of defining the risk certificate in absolute terms, we normalise the risk certificate such that $\tilde{\alpha}=0\%$ corresponds to bidding the point-forecast and $\tilde{\alpha}=100\%$ to the all-or-nothing solution. The trading strategies are examined for a range of risk certificates from $0\%$ to $100\%$ with increments of $25\%$. Two cases are compared to evaluate the trading performance: (1) the \textit{no price-impact} case, where the historical balancing prices remain constant regardless of the trading strategy and (2) the \textit{price-impact} case, where the impact of the trading strategy on the balancing price is assessed using the method described in Subsection \ref{subsec:price-impact}. Given that we consider a large-scale offshore wind farm, it is important to determine whether the trading strategy would have substantial price-impact. The comparative analysis examines how the historical evaluation would be affected when the price-impact is not considered.


Probabilistic forecasts of wind power generation are obtained with the Incremental Quantile Functions model proposed by \cite{park2022learning}. The model was established by combining weather forecast data with production data of the offshore wind farm for the year 2023. The conditional mean $\E\left[E_t\right]$ and variance $\Var\left(E_t\right)$ of the energy production are computed using this forecasting model. We consider the observed day-ahead and balancing prices as inputs rather than using electricity price forecasts. Although price forecasts are required in reality, this is a different research area that falls outside of the scope of this paper. The results can therefore be interpreted as a theoretical upper bound on the profit of the trading strategy. The balancing prices are first resampled from a 15-minute to a 1-hour resolution using the mean value. Resampling is required because the day-ahead market and the balancing market are not on the same time resolution, as explained in Appendix \ref{app:full_problem}.

Short-term electricity market data for Belgium is obtained from The Open Data Platform by Elia \cite{eliaopendata}, the Belgian transmission system operator, and the Transparency Platform by the European Network For Transmission System Operators for Electricity \cite{entsoe}. The individual balancing volume bids (see Section \ref{subsec:price-impact}) are also obtained from Elia. Moreover, we use representative data from the SCADA system of a wind farm in the Belgian offshore zone and weather forecasts from well-known Numerical Weather Prediction providers such as Deutscher Wetterdienst and Méteo France.

\subsection{Price-impact evaluation}\label{subsec:price-impact}

To evaluate the price-impact ex-post, we first need to project the impact of the trading strategy on the system imbalance and determine the required balancing volume, i.e., the opposite of the system imbalance. Afterwards, the balancing price can be calculated from the historical balancing volume bids.

The system imbalance resulting from the trading strategy $\psi^{\alpha}_t$ is given by:
\begin{equation}
    \psi^{\alpha}_t = \psi_t + 4\left(E_t-y_t\right),
\end{equation}
with $\psi_t$ the historical system imbalance over quarter-hour $t$. Since the historical system imbalance is in MW, whereas the production and day-ahead bid are in MWh over a 15-minute time window, the open position is multiplied by a factor of 4.

\begin{figure}[t!]
    \centering
    \includegraphics[width=0.42\textwidth]{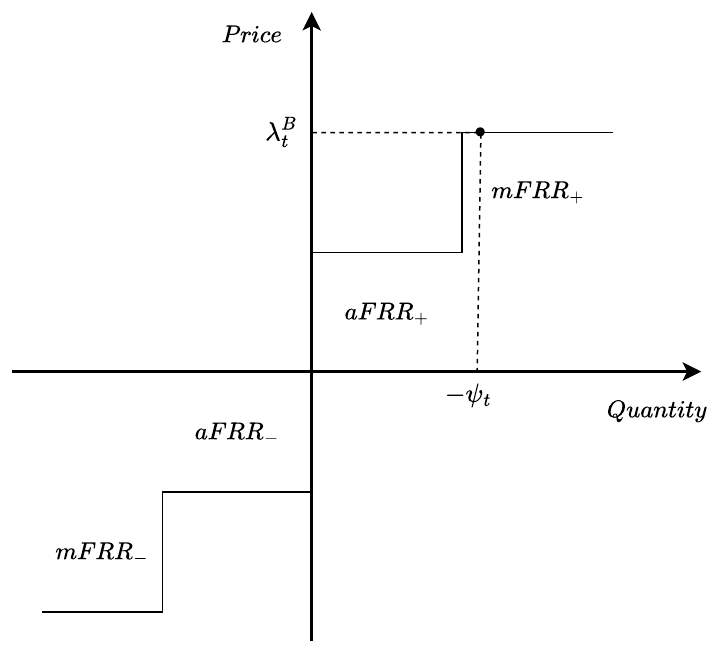}
    \caption{A simplified example of the merit-order curve and balancing price calculation in case of negative system imbalance $\psi_t<0$. The incremental ($aFRR_+$ and $mFRR_+$) and decremental ($aFRR_-$, $mFRR_-$) balancing volumes are sorted based on marginal price and activation order. The balancing price $\lambda^{\textrm{B}}_t$ is determined by the required balancing volume, i.e., the opposite of the system imbalance $-\psi_t$. Prices are in $EUR/MWh$ and quantities in $MW$.}    
    \label{fig:merit_order}
\end{figure}

For the balancing price calculation, we first derive the merit-order curves using historical balancing volume bids. To this end, we sort the balancing volumes on activation order and marginal price. The activation order requires that the automatic Frequency Restoration Reserve (aFRR) volumes are always activated before the manual Frequency Restoration Reserve (mFRR) volumes. After considering the activation order, the bids are sorted on marginal price to ensure that the cheapest volumes are used first. The balancing price can then be determined by the system imbalance, selecting the bid that covers the required volumes to restore the balance. The methodology is also illustrated in Figure \ref{fig:merit_order}. 

Note that certain simplifications were made concerning the balancing price calculation. First, the correction factor for large system imbalances by Elia \cite{elia2024balancing} is not considered, as it only applies to the Belgian setting. Additionally, the balancing price is determined using the average system imbalance over the settlement period instead of the instantaneous system imbalance and the effect of imbalance netting is disregarded. Last, we assume that balancing volumes are not shared across borders, which will change and has already changed recently due to the MARI \cite{entsoe_mari} and PICASSO \cite{entsoe_picasso} projects. To ensure comparability, the historical balancing prices are therefore recalculated according to the presented methodology.   

\subsection{Numerical results and discussion}

\begin{figure*}[h!]
    \vspace{22pt}
    \centering
    \begin{subfigure}{0.48\textwidth}
        \includegraphics[width=\textwidth]{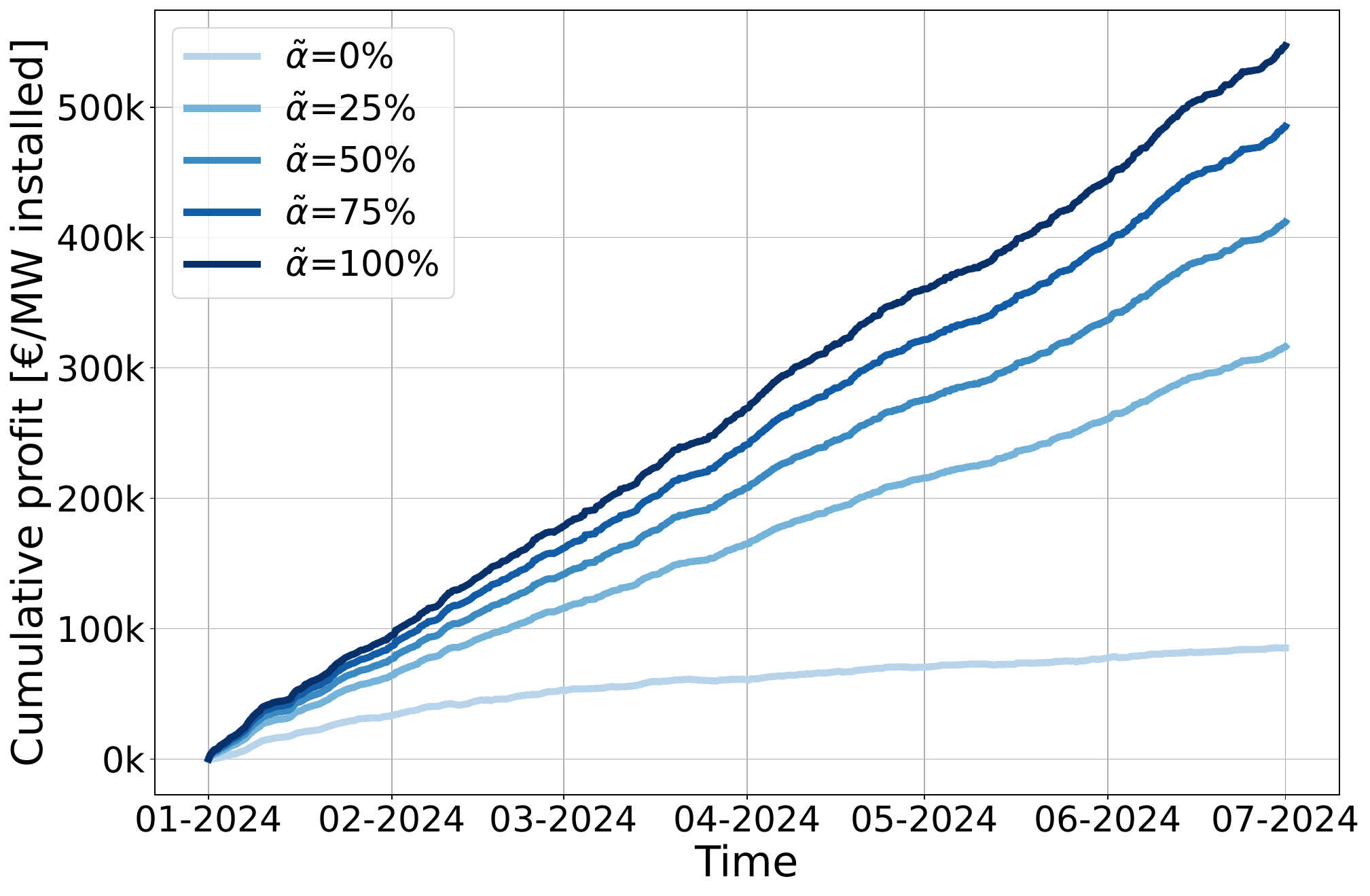}
        \subcaption{No price-impact}\label{subfig:backtest-pt}
    \end{subfigure}
    \begin{subfigure}{0.48\textwidth}
        \includegraphics[width=\textwidth]{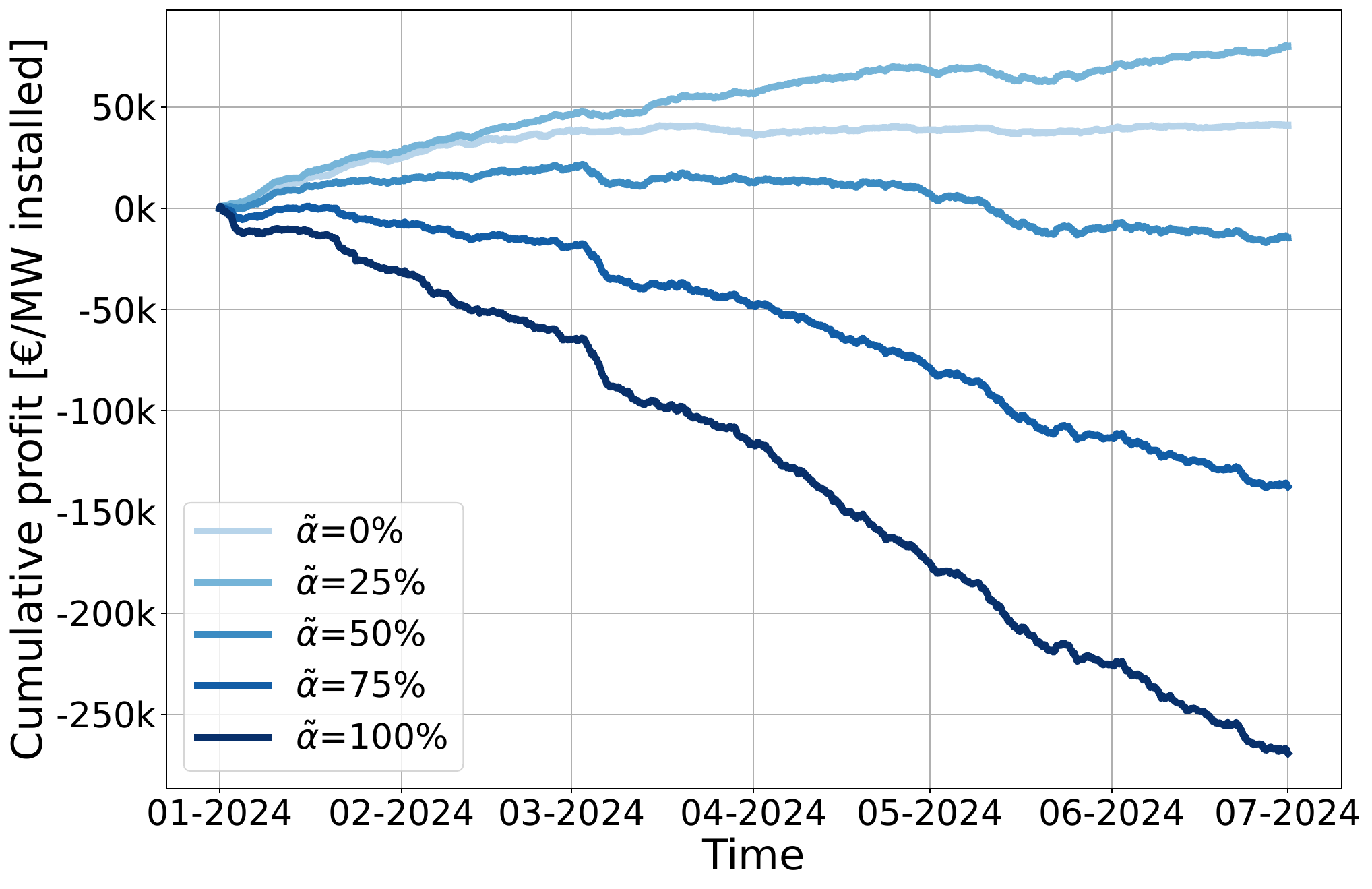}
        \subcaption{Price-impact in balancing market}\label{subfig:backtest-pm}
    \end{subfigure}
    \caption{Cumulative profit of different trading strategies defined based on the risk certificate $\tilde{\alpha}\in[0\%,25\%,50\%, 75\%, 100\%]$. Evaluation is done using historical data of the first semester of 2024. Figure (a) illustrates the trading performance in case the balancing price remains constant regardless of the trading strategy. Figure (b) shows how the results change when the price-impact is considered during evaluation. Cumulative profit is normalised by the installed capacity of the wind farm.}
    \vspace{25pt}
    \label{fig:backtest}
\end{figure*}


\begin{figure*}[h!]
\centering
\begin{tabular}{cccc}
\includegraphics[width=0.3\textwidth]{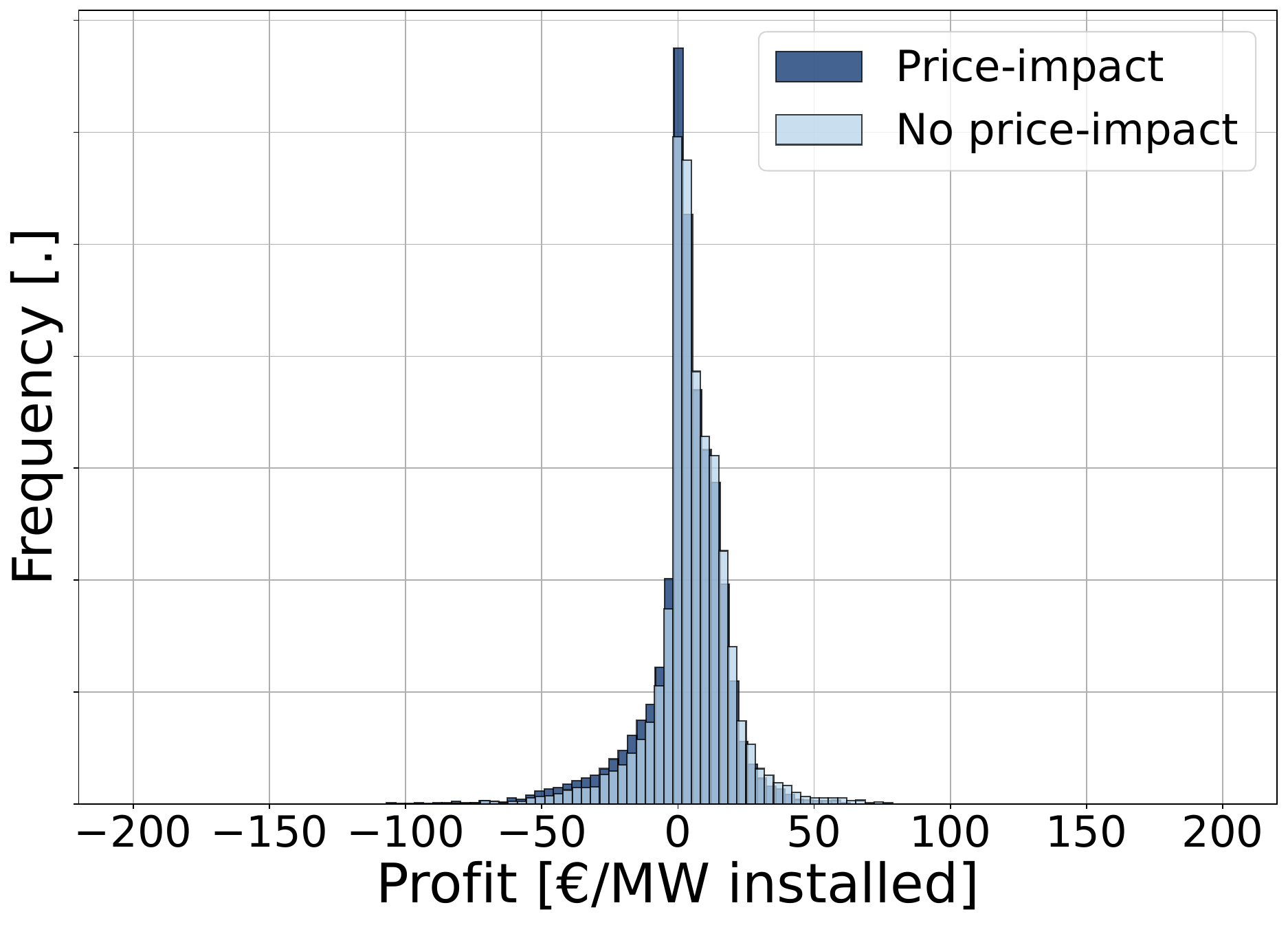} &
\includegraphics[width=0.3\textwidth]{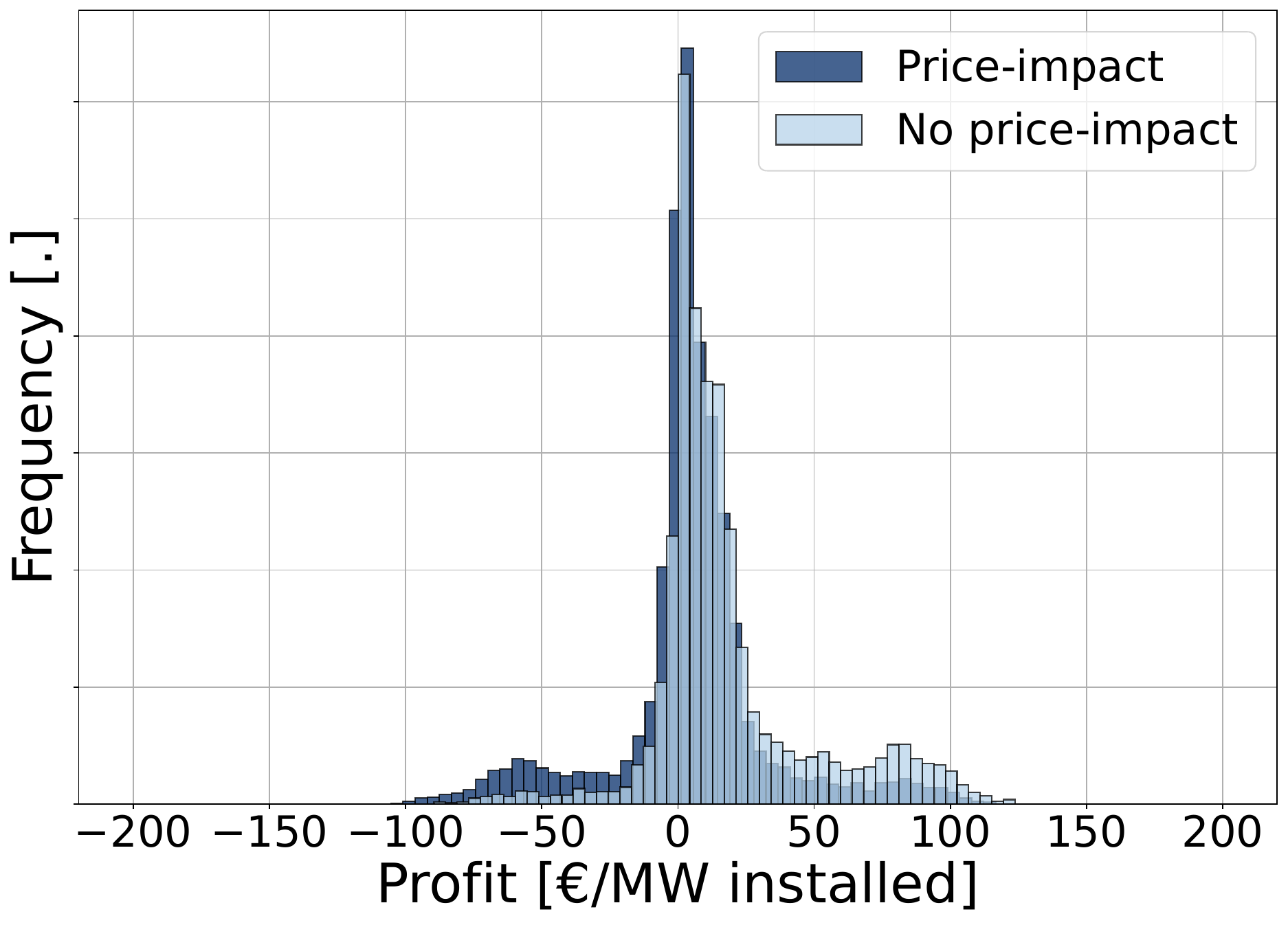} &
\includegraphics[width=0.3\textwidth]{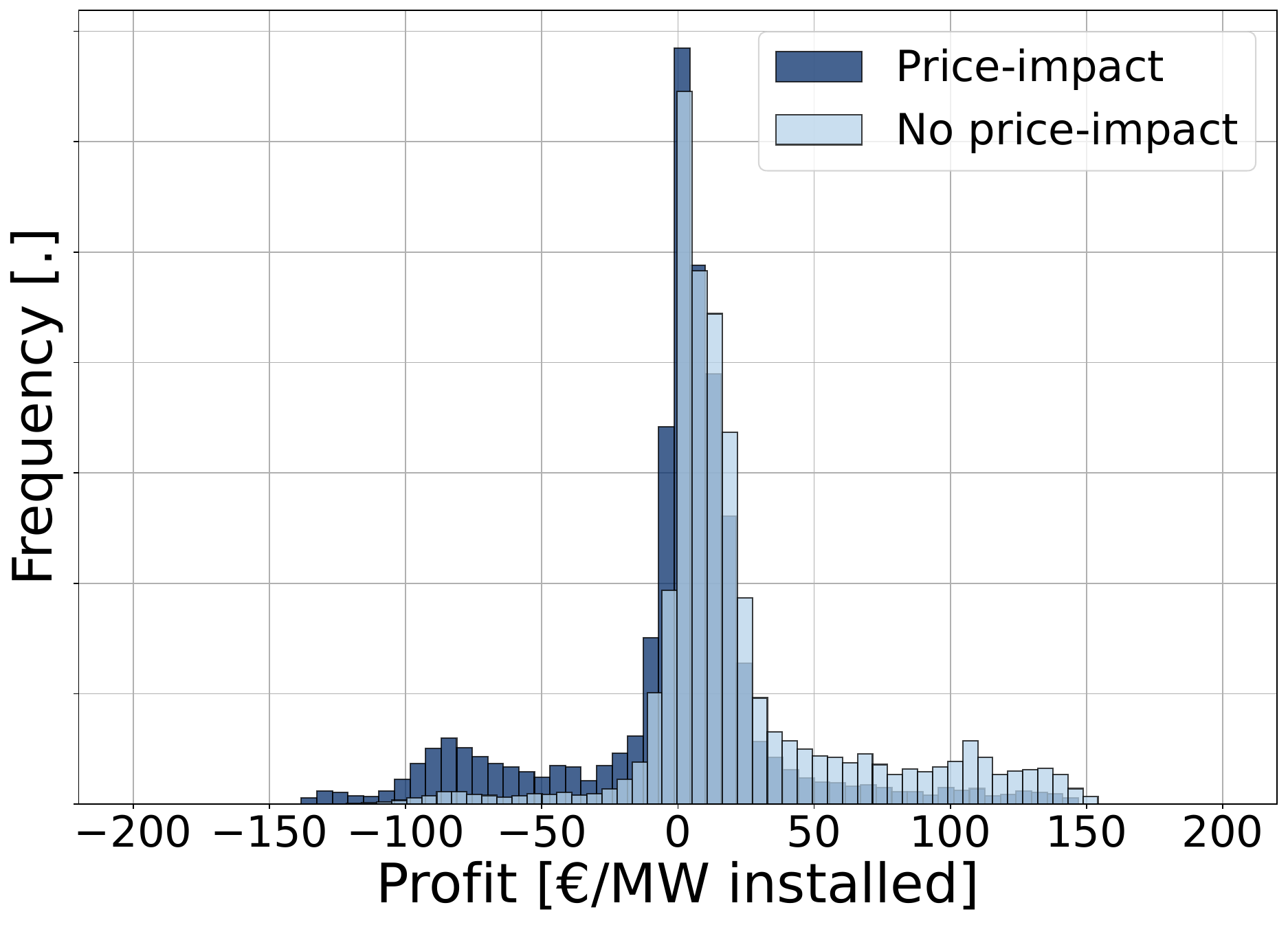} \\
\textbf{(a) $\tilde{\alpha}=0\%$}  & \textbf{(b) $\tilde{\alpha}=25\%$} & \textbf{(c) $\tilde{\alpha}=50\%$}  \\[6pt]
\end{tabular}
\begin{tabular}{cccc}
\includegraphics[width=0.3\textwidth]{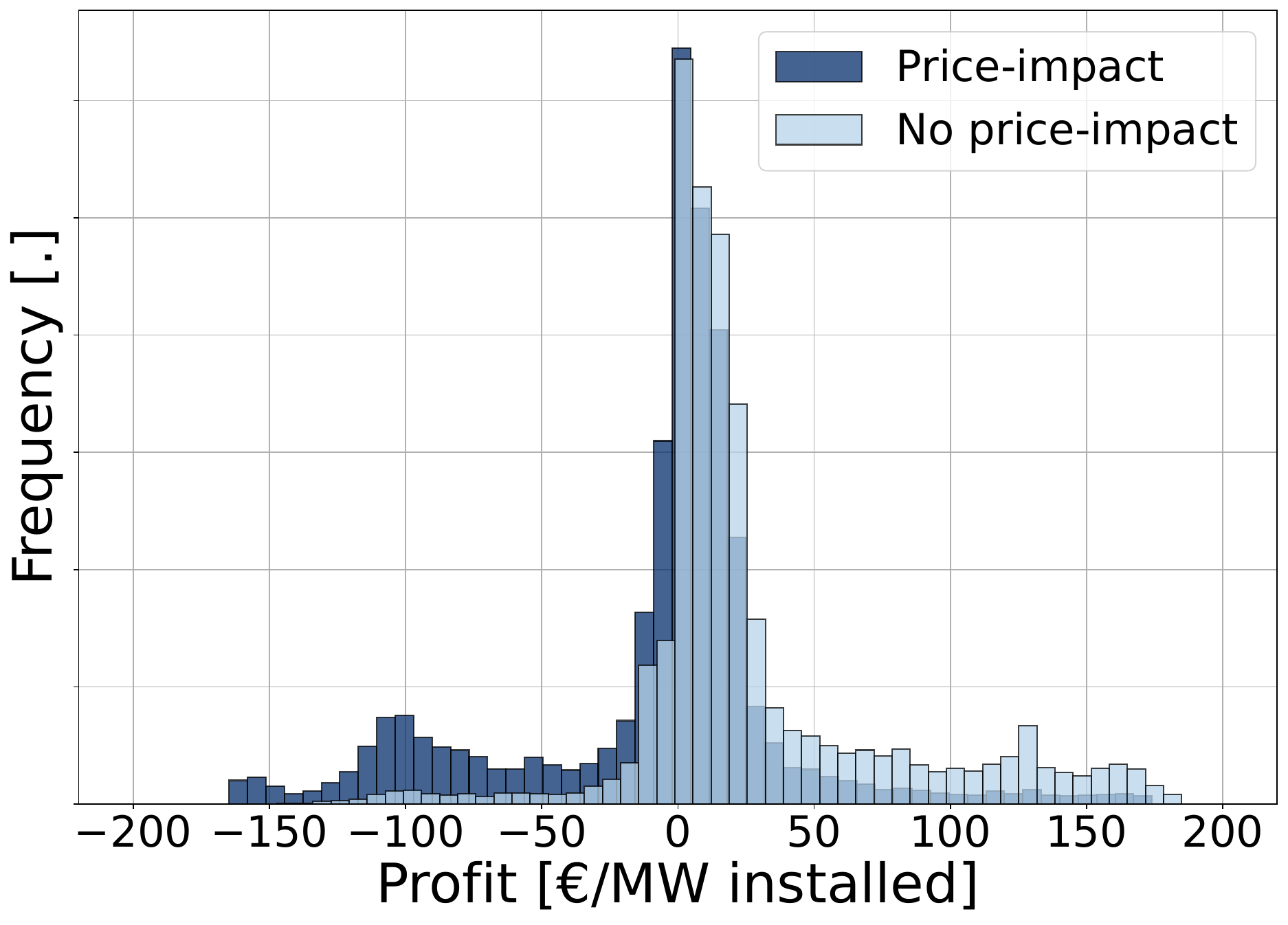} &
\includegraphics[width=0.3\textwidth]{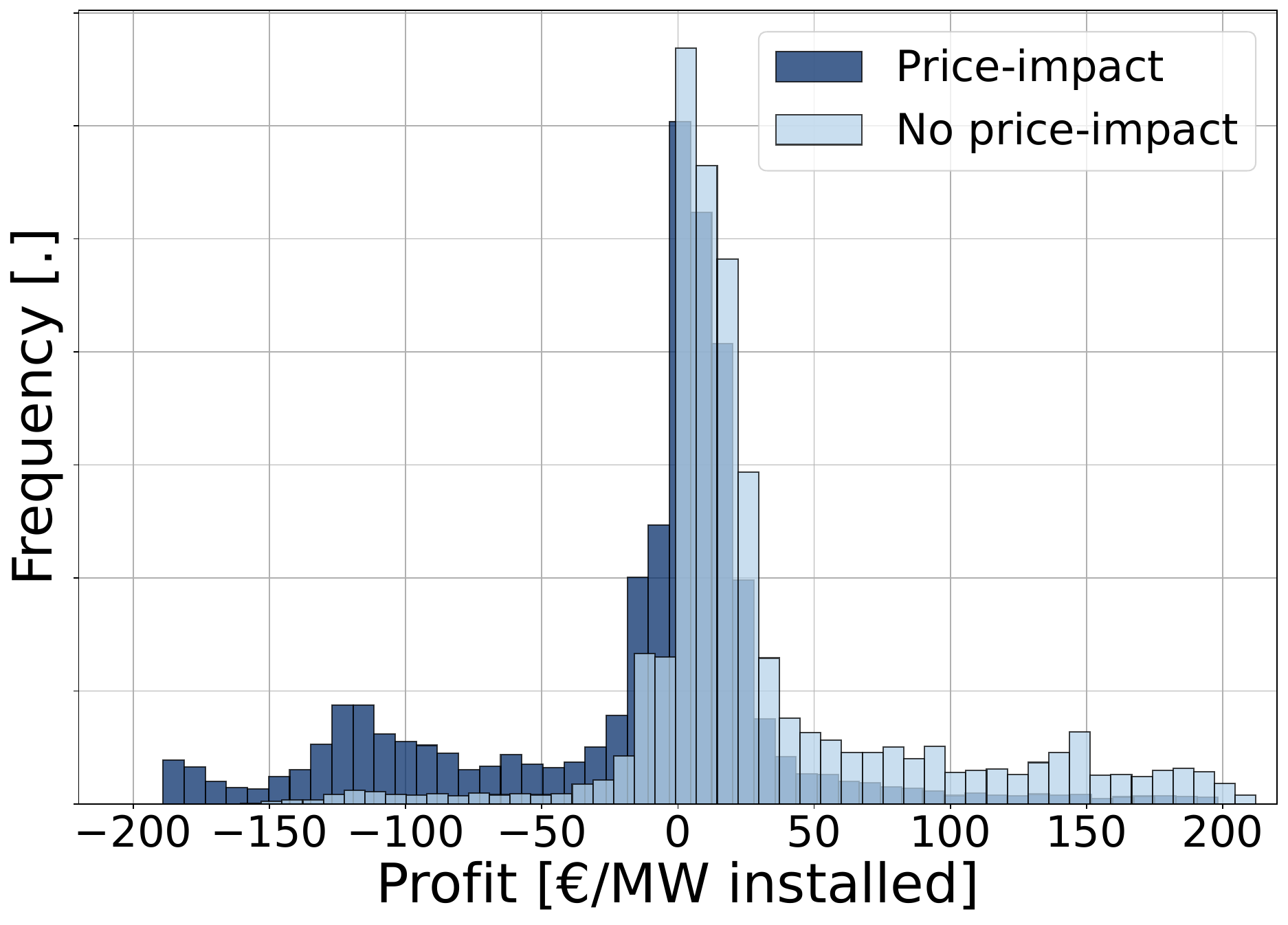} \\
\textbf{(d) $\tilde{\alpha}=75\%$}  & \textbf{(e) $\tilde{\alpha}=100\%$}  \\[6pt]
\end{tabular}
\caption{Comparison of the profit distributions of the trading strategies between the case where the price-impact is not considered (light blue) and the case where the price-impact on the balancing market is considered (dark blue). Different trading strategies are defined based on the risk certificate $\tilde{\alpha}\in[0\%,25\%,50\%, 75\%, 100\%]$. Profit is normalised by the installed capacity of the wind farm.}
\vspace{22pt}
\label{fig:profit_dist}
\end{figure*}

Figure \ref{fig:backtest} shows the cumulative profit evaluated on historical data for the two cases. The cumulative profit when the price-impact is not considered during ex-post evaluation can be found in Figure \ref{subfig:backtest-pt}. In this case, the empirical results are consistent with the objective of the optimisation problem, as allowing for larger positions in the balancing market reflects higher returns. Note that a wind farm with a modest installed capacity, for instance a small-scale onshore wind farm, would observe these results. Figure \ref{subfig:backtest-pm} shows the cumulative profit in case the price-impact on the balancing market is considered. We find that the all-or-nothing strategy ($\tilde{\alpha}=100\%$) is not optimal anymore and even results in substantial long-term losses. Nonetheless, our risk-constrained method ($\tilde{\alpha}=25\%$) can still outperform the traditional strategy to bid the point-forecast ($\tilde{\alpha}=0\%$) in this case.

Figure \ref{fig:profit_dist} shows how the profit distribution for different trading strategies compare in the case with and without price-impact. When the day-ahead bids are equal to the point-forecast ($\tilde{\alpha}=0\%$), we find that the profit distributions are comparable. However, the profit distributions start to shift when we increase the risk certificate $\tilde{\alpha}$. The case without price-impact shows more probability mass in the right tail, indicating potential gains, and less probability mass in the left tail, indicating potential losses. The effect becomes more profound as the trading strategy allows for larger open positions in the balancing market.

The findings illustrate the importance of determining the price-impact on the balancing market when evaluating trading strategies on historical data. When the price-impact is not considered, the evaluation does not only result in overconfidence about the performance of the trading strategy in the long-run, but also in a critical underestimation of the short-term risks. In addition, we find that the profit distribution is dependent on the trading strategy through the balancing price. This implies that, when historical balancing prices are used for risk measurement, the true risk of the trading strategy will not be reflected by the chosen risk measure.

The observations can be explained by considering the historical system imbalance and installed capacity of the wind farm. In the Belgian offshore zone, the installed capacity of a farm is certainly larger than 100 MW, whereas the historical system imbalance has been between -100 MW and 100 MW in almost 58\% of the cases over the evaluation period. Hence, a bidding strategy that correctly anticipates the electricity prices can cause the system imbalance to switch direction, making the potential balancing revenue become a penalty. 

\section{Conclusion}
\label{sec:conclusion}

This paper investigates day-ahead bidding strategies from the perspective of the wind farm operator. We argue that this is a decision problem under uncertainty, where the objective is to maximise expected operational profit while limiting the risk reflected by open positions in the balancing market. A stochastic optimisation problem is presented to represent the risk-return trade-off, and we derive an analytical solution to this problem. The simple analytical solution allows us to study the problem in more detail by evaluating historical trading performance. To this end, we show how the price-impact on the balancing market can be considered in the ex-post evaluation and compare this to the case where we do not consider this impact.

The results indicate that, taking into account the price-impact on the balancing market, the all-or-nothing strategy is not optimal and results in serious financial losses in the long-run. The fact that the all-or-nothing strategy is sub-optimal can be explained by the installed capacity of an offshore wind farm, which is sufficiently large to impact the balancing price. However, our risk-constrained method can still increase operational profit in the long-run compared to traditional point-forecast bidding. Furthermore, the findings illustrate that, the profit distribution is dependent on the trading strategy through the balancing price. Indicating that risk measurement using historical balancing prices could significantly underestimate the financial losses. Therefore, the analysis reveals the importance of considering the price-impact while evaluating trading strategies on historical data for large-scale offshore wind farms. 

An important direction for future research is to incorporate the price-impact on the balancing market in the optimisation problem. Nonetheless, this significantly complicates the problem to a bi-level stochastic optimisation problem with an unknown merit-order curve. \cite{zugno2013pool} has shown how to solve this problem using a mathematical program with equilibrium constraints. However, the authors used a simplified scenario-based merit-order curve to obtain their solution, not necessarily reflecting the real-world. Furthermore, the computational complexity of the method makes it infeasible to construct an adaptive trading strategy. An alternative approach would be to learn the optimal solution directly from data through reinforcement learning. Reinforcement learning derives the optimal strategy by interacting with the environment and adjusting its strategy based on the outcome. Several works in the field of electricity markets considering reinforcement learning have shown promising results \cite{harder2023finding,harder2023fit,miskiw2024multi}, but, to the best of our knowledge, it has not yet been applied to day-ahead bidding for renewable energy producers with price-impact on the balancing market.

\section{Acknowledgments}

The authors gratefully acknowledge the financial support of the Flemish Government through the Flanders AI Research Program, the Energy Transition Funds of the Belgian Federal Government through the BeFORECAST project and the Sustainable Blue Economy Partnership through the INSPIRE project (project no. SBEP2023-440). The first author would also like to thank Rob Coutuer and Vic Peeters for their insightful discussions on the subject.

\printbibliography

\appendix

\section{Appendix}\label{app:full_problem}

The problem statement, in Section \ref{sec:problem}, simplified the formulation by considering the resolution of the day-ahead and balancing market to be equivalent $\mathcal{W}^{\textrm{DA}} = \mathcal{W}^{\textrm{B}}$. However, these resolutions are not always the same, for instance, in Belgium the day-ahead market has a resolution of 1-hour and imbalances are settled on a 15-minute basis \cite{eliaimbalance}.\\
\\
Therefore, we consider the case where the resolution of the imbalance settlements is more granular $r^{\textrm{B}}<r^{\textrm{DA}}$ but still a multiple of the day-ahead resolution $n r^{\textrm{B}} = r^{\textrm{DA}}$ with $n\in\mathbb{Z}$.\\
\\
The objective function (\ref{maxi:objective}) can now be rewritten as:
\begin{equation}\label{eq:obj_full}
    \E\left[\Lambda^{\textrm{DA}}_t y_t + \sum_{\substack{t'\\w^{\textrm{B}}_{t'} \in \mathcal{W}^{\textrm{B}}\\w^{\textrm{B}}_{t'} \subset w^{\textrm{DA}}_t}} \Lambda^{\textrm{B}}_{t'}(E_{t'}-y_{t'})\right].
\end{equation}\\ 
\\
Because of a different time resolution in the day-ahead and the balancing market, the day-ahead bidding quantity $y_t$ results in a balancing obligation $y_{t'}$ of:
\begin{equation}\label{eq:balancing_obligation}
    y_{t'} = \frac{1}{n}y_t \quad \quad \forall w^{\textrm{B}}_{t'}\in \mathcal{W}^{\textrm{B}}, w^{\textrm{B}}_{t'} \subset w^{\textrm{DA}}_{t},
\end{equation}
with $y_u$ in MWh for $u\in\{t,t'\}$.\\
\\
Substituting (\ref{eq:balancing_obligation}) in (\ref{eq:obj_full}) gives:
\begin{equation}
    \E\left[(\Lambda^{\textrm{DA}}_t - \Lambda^{\textrm{B}}_t) y_t + C\right],
\end{equation}
with:
\begin{equation}\label{eq:resample_balancing_price}
    \Lambda^{\textrm{B}}_t = \frac{1}{n}\sum_{\substack{t'\\w^{\textrm{B}}_{t'} \in \mathbb{W}^{\textrm{B}}\\w^{\textrm{B}}_{t'} \subset w^{\textrm{DA}}_t}}\Lambda^{\textrm{B}}_{t'},
\end{equation}
\begin{equation}
    C = \sum_{\substack{t'\\w^{\textrm{B}}_{t'} \in \mathbb{W}^{\textrm{B}}\\w^{\textrm{B}}_{t'} \subset w^{\textrm{DA}}_t}}\Lambda^{\textrm{B}}_{t'}E_{t'}.
\end{equation}\\
\\
Therefore, we find that the objective function is again linear in $y_t$:
\begin{equation}
    \left(\E\left[\Lambda^{\textrm{DA}}_t\right] - \E\left[\Lambda^{\textrm{B}}_t\right]\right)y_t + \E[C].
\end{equation}\\
\\
Following the reasoning in Section \ref{sec:analytical}, the analytical solution is given by:
\begin{equation}
    y^{*}_t = 
    \begin{cases}
        \min\{\E\left[E_t\right]+\Delta_t ; \beta\} & \text{if } \E\left[\Lambda^{\textrm{DA}}_t\right]>\E\left[\Lambda^{\textrm{B}}_t\right]\\
        \max\{\E\left[E_t\right]-\Delta_t ; 0\} & \text{else},
    \end{cases}
\end{equation}
with $\Delta_t = \sqrt{\alpha-\Var\left(E_t\right)}$.

\end{document}